\documentclass[12pt]{iopart}

\usepackage[dvips]{graphicx}
\usepackage{dcolumn}
\usepackage{bm}
\usepackage{iopams}
\usepackage{graphicx}
\begin{document}
\title{Two-dimensional array of microtraps with atomic shift register on a chip}
\author{S Whitlock, R Gerritsma\footnote{Present address: Institut f\"{u}r Quantenoptik und Quanteninformation der \"{O}sterreichischen Akademie der Wissenschaften, Innsbruck, Austria}, T Fernholz\footnote{Present address: Midlands Ultracold Atom Research Centre, School of Physics \& Astronomy, The University of Nottingham, Nottingham NG7 2RD, UK} and R J C Spreeuw}
\address{Van der Waals-Zeeman Institute, University of
Amsterdam, Amsterdam, The Netherlands.}
\ead{S.M.Whitlock@uva.nl}

\date{\today}


\begin{abstract}
Arrays of trapped atoms are the ideal starting point for developing registers comprising large numbers of physical qubits for storing and processing quantum information. One very promising approach involves neutral atom traps produced on microfabricated devices known as atom chips, as almost arbitrary trap configurations can be realised in a robust and compact package. Until now, however, atom chip experiments have focused on small systems incorporating single or only a few individual traps. Here we report experiments on a two-dimensional array of trapped ultracold atom clouds prepared using a simple magnetic-film atom chip. We are able to load atoms into hundreds of tightly confining and optically resolved array sites. We then cool the individual atom clouds in parallel to the critical temperature required for quantum degeneracy. Atoms are shuttled across the chip surface utilising the atom chip as an atomic shift register and local manipulation of atoms is implemented using a focused laser to rapidly empty individual traps.\end{abstract}

\section{Introduction}
Of central importance in the quest for a quantum computer is the issue of scalability. In the \emph{trapped-ion} approach for example, microfabricated-chip traps~\cite{CirZol00,KieMonWin02} have recently been able to accommodate several ions~\cite{SeiChiWin06,StiHenMon06} and shuttle them between different zones~\cite{HenOlmRab06}. \emph{Neutral atoms} on the other hand~\cite{DeuBrePou00} possess the significant advantage of an intrinsically weaker interaction with the environment, yet are routinely manipulated using optical and magnetic trapping techniques. So far, the primary tool used to create registers of neutral atoms has been the optical lattice, which allows near-perfect arrays of atoms with typical separations of half the optical wavelength~\cite{GreManBlo02}; however addressing and locally manipulating atoms on this length-scale is a serious challenge. This has motivated the development and use of large-period optical lattices~\cite{SchDotMes04,NelLiWei07} or patterned loading techniques~\cite{PeiPorPhi03}, as well as addressable arrays of optical traps produced using holographic techniques~\cite{BerDarGra04} or micro-lens arrays~\cite{DumVolErt02}.

Magnetic microtraps integrated on atom chips~\cite{HanHomRei01,OttForZim01,Rei02,FolKruHen02} are a promising alternative to these optical approaches. An atom chip typically consists of a microscopic pattern of wires (micro-electromagnets) or permanent magnets~\cite{SinCurHin05,HalWhiSid06,FerGerSpr08} arranged on a substrate as to produce tailor-made magnetic potentials for ultracold atoms. Qubits encoded in the ground-state hyperfine levels of the trapped atoms could offer exceptionally long coherence times~\cite{TreHomRei04}, and quantum-logic gates could be implemented via controlled collisions~\cite{CalHinZol00} or by long-range interactions using laser excited Rydberg states~\cite{JakCirLuk00}. With established micro/nano-fabrication techniques it is then feasible to scale-up and integrate many microtraps on a single chip. In earlier work, one-dimensional periodic potentials~\cite{GunKraFor05,SinCurHin05,BoyStrPri07,ManVolHan08} and small arrays consisting of several traps~\cite{HanHomRei01,GraPfa03} have been produced using both current carrying wires and magnetic materials. The present work greatly extends on these results to create and study a two-dimensional array of traps accommodating more than $500$ atom clouds. This is made possible using two-dimensional structures lithographically patterned into magnetic films~\cite{GerWhiSpr07} which allow the integration of hundreds of tightly confining traps in custom-designed and optically resolvable microtrap arrays.

\begin{figure} \centering\includegraphics[width=0.6\columnwidth]{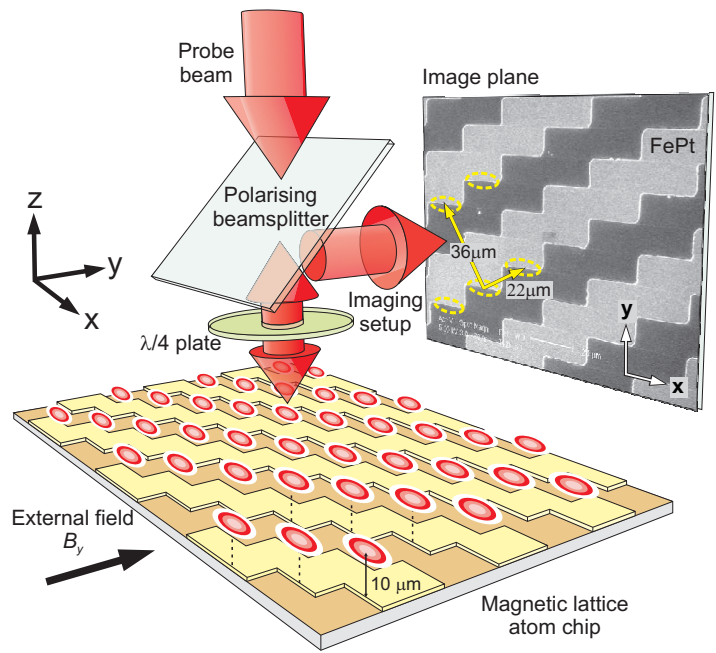}
\caption{{\bf Magnetic lattice atom chip and experimental setup.} The experiments are performed using a FePt magnetic film, lithographically patterned with a two-dimensional lattice with periods of $22~\mu$m and $36~\mu$m and coated with a reflective gold layer.  The film is uniformly magnetised along $z$ and combined with an external field $B_y$ to create an array of magnetic microtraps for atoms positioned $10~\mu$m from the film surface. For detection, the trapped atoms are illuminated by a retro-reflected probe laser beam which is partly absorbed by the atoms and imaged onto a CCD camera for analysis. Projected on the backplane of the figure is a microscope image of the FePt film with a few trap positions indicated by yellow ellipses.} \label{fig:setup}
\end{figure}

\begin{figure*} \centering\includegraphics[width=1\columnwidth]{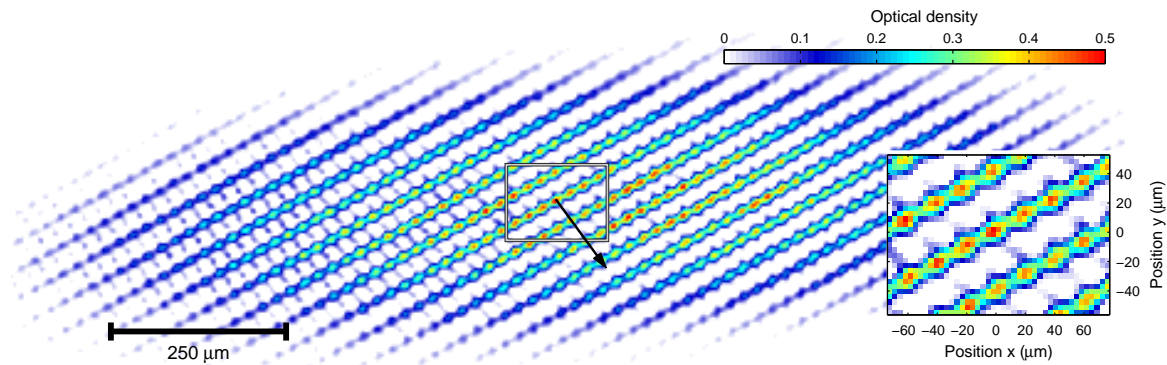}
\caption{{\bf Loaded array of magnetic microtraps.} An absorption image of the two-dimensional atomic distribution (shown in false-colour) after transfer to the magnetic lattice. The optical resolution is approximately $7~\mu$m. The inset shows a magnified region of the central part of the array as indicated by the outlined box. Atoms populate more than 500 array sites over a region of $1.4\times0.4$~mm$^2$ of the chip surface; the scale bar corresponds to a distance of $250~\mu$m. The solid arrow indicates the direction of transport for the atomic shift register experiment (figure~\ref{fig:shiftregister}).} \label{fig:array}
\end{figure*}

\section{Magnetic microtrap arrays}
At the core of our experiment (figure~\ref{fig:setup}) is an atom chip~(for design and fabrication details see ref.~\cite{GerWhiSpr07}), which incorporates a single 300-nm-thick FePt film, patterned using optical lithography, then gold-coated and magnetised in the direction perpendicular to the film surface ($M_z=\nobreak670$~kA/m). Combined with an external field ($B_y\!=\!\nobreak22.5$~G), this simple setup produces an array of magnetic field minima (magnetic lattice), which is used to trap ultracold $^{87}$Rb atoms in the $|F\!\!=\nobreak\!\!2,m_F\!\!=\nobreak\!\!2\rangle$ state.  In the following experiments, each trap is produced $10~\mu$m from the surface and provides tight harmonic confinement to the atoms (with trapping frequencies $\omega_\parallel/2\pi=\nobreak5.5$~kHz and $\omega_\perp/2\pi=\nobreak13.4$~kHz in the axial and transverse directions respectively). The distances between neighbouring traps are $22~\mu$m and $36~\mu$m (1250~traps/mm$^2$), chosen to facilitate optical addressing. The barriers between traps, the trapping frequencies and the trap positions are precisely controlled in parallel through the external field strength and orientation. We believe magnetic-film atom chips are best suited for high-density integration, as for example, an equivalent potential produced using currents through micron-sized gold wires would require a power dissipation of $\sim$30~W/mm$^2$.

We first describe the loading of the microtrap array. A cloud of $^{87}$Rb atoms is collected and cooled in a mirror-magneto-optical-trap and then transferred to a single magnetic trap created by a current-carrying wire and the external field~\cite{Rei02}. A two-dimensional view of the atomic distribution is obtained using reflective resonant absorption imaging (figure~\ref{fig:setup}). The cloud contains $N=1\times10^6$ atoms at a temperature of $\sim30~\mu$K before it is transferred to the film surface by ramping off the wire current. The atom cloud first becomes corrugated as the trap merges with the field of the lattice, until for zero wire current the atoms are confined in the microtrap array alone (figure~\ref{fig:array}). We determine the number of atoms in each trap by integrating the absorption over each unit cell. During merging, the atom cloud expands over an area of $\sim\!1.4\times0.4$~mm$^2$, populating more than 500 traps, each with $N\!>\!200$ and up to 2500 atoms, showing it is possible to transfer atoms from a single magnetic trap to hundreds of microtraps. After loading, we observe atom loss over two characteristic time-scales: minor initial loss over the first 100~ms, caused by residual evaporation of atoms over the potential barriers, then background loss, occurring with a time-constant of around 2~s.

\begin{figure*}[t] \centering\includegraphics[width=1\columnwidth]{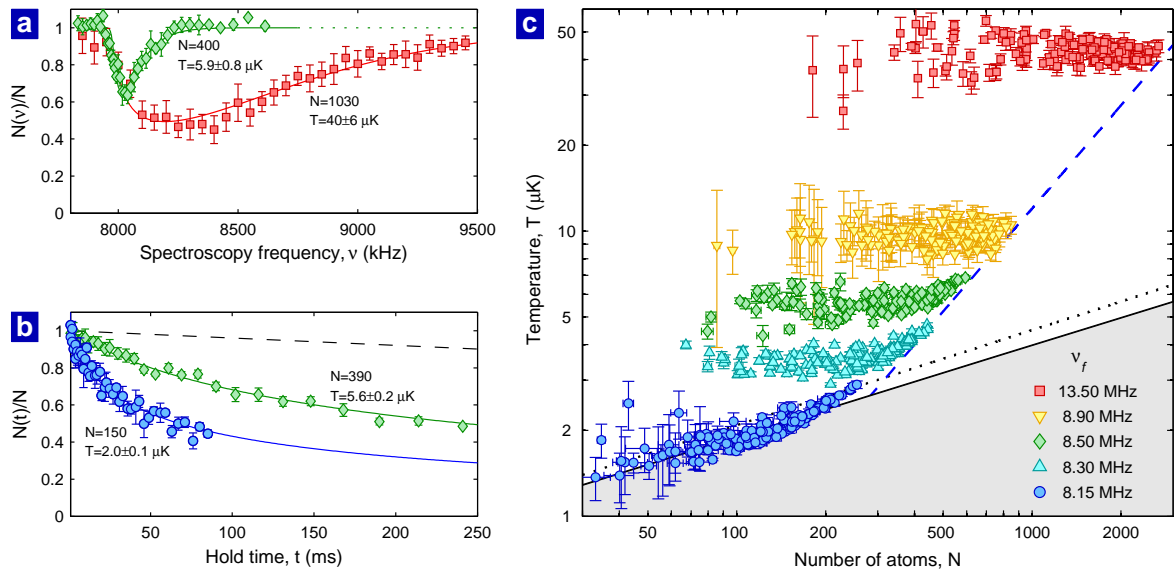}
\caption{{\bf Radio-frequency evaporative cooling and site-resolved thermometry.} Forced radio-frequency (rf) evaporation is used to cool the individual atom clouds to the critical temperature for quantum degeneracy. Site-resolved rf spectroscopy and three-body decay measurements are performed to determine the distribution of cloud temperatures. {\bf (a)} Two representative rf spectra and fits (solid lines) for one selected site for final evaporation frequencies of $\nu_f=13.5$~MHz (red squares) and $\nu_f=8.50$~MHz (green diamonds), with fitted values for $N$ and $T$ labelled.The fraction of atoms remaining in a well after the application of an rf outcoupling pulse is plotted along the vertical axis. {\bf (b)} Two representative three-body decay curves and fits (solid lines) for the same array site as in (a) for $\nu_f=8.50$~MHz and $\nu_f=8.15$~MHz (blue circles). The dashed line indicates the expected loss due to background vapour collisions with $\tau=2$~s. {\bf (c)} The full set of data $(N,T)$, for $>\!150$ array sites and five different final evaporation frequencies. The error-bars represent statistical uncertainties estimated from several repetitions of the experiment. The blue dashed line indicates a constant evaporation efficiency of $d(\ln T)/d(\ln N)=1.2$ calculated for one site. The dotted line marks the Bose-Einstein condensation transition (degeneracy temperature) in the thermodynamic limit neglecting interactions. The numerically calculated boundary to the grey region is for a constant phase-space-density of $\rho_0=2.612$, corresponding to the onset of Bose-Einstein condensation including the effects of finite atom number and interactions.} \label{fig:evaporation}
\end{figure*}

\section{Evaporative cooling and site-resolved thermometry}
\subsection{Radio frequency evaporation and spectroscopy}
For storing quantum information, it is desired to initialise the register to a well-defined state with negligible thermal excitations. One powerful feature of magnetic lattices is the availability of radio-frequency cooling techniques, which we show can be applied to cool hundreds of atom clouds in parallel. We perform forced evaporative cooling using a two-part radio-frequency (rf) sweep. Radio-frequency spectroscopy~\cite{FerGerSpr08} is then used to characterise more than 150 microtraps near the center of the array (figure~\ref{fig:evaporation}).

Sweeping the radio frequency from high to low frequency continuously couples the more energetic atoms to untrapped Zeeman states (which subsequently leave the traps) while the remaining atoms rethermalise to lower temperatures. In our experiments the field minima of the potentials is around $11.41\pm0.05$~G corresponding to a Zeeman energy splitting of $\sim7.99$~MHz. We first ramp from 13.5~MHz to 8.5~MHz in 200~ms (corresponding to effective trap depths of around $48~\mu$K), then from 8.5~MHz to 8.15~MHz in 35~ms ($15~\mu$K). In rf-spectroscopy, an rf field is pulsed on for $\sim1$~ms with a variable frequency to selectively outcouple atoms depending on their energy. We record an absorption image to determine the number of atoms remaining in each trap after the pulse and repeat the experimental cycle for each value of the spectroscopy-frequency to construct a collection of energy distributions (one for each site~(figure~\ref{fig:evaporation}a)). The amplitude of the rf field is adjusted for each dataset to ensure a sufficient number of atoms are outcoupled for detection (typically 30\% to 50\% at peak), with typical Rabi frequencies of $\Omega\approx 2\pi\times15~$kHz and $\Omega\approx 2\pi\times4~$kHz for $\nu_f=13.50$~MHz and $\nu_f=8.50$~MHz datasets respectively. We fit the measured spectra to a spatial outcoupling model~\cite{FerGerSpr08} based on the calculated atomic distributions (see below) and including finite spectral resolution, to determine the cloud temperatures and the magnetic field strengths at the trap minima. For the spectra shown in figure~\ref{fig:evaporation}a we observe low energy tails associated with broadening widths of 30~kHz and 10~kHz for $\nu_f=13.50$~MHz and $\nu_f=8.50$~MHz datasets respectively. 

Radio-frequency spectroscopy is performed at several intervals during the evaporation sweep, and analysed independently for each microtrap to determine the cloud temperature $T$ and the magnetic field strength at the trap minimum (figure~\ref{fig:evaporation}a). Before evaporation, we find the cloud temperatures are nearly equal for all traps (red-squares in figure~\ref{fig:evaporation}c) with $\langle T\rangle=43\pm4(3)~\mu$K (most-likely constrained by the trap depth of around $400~\mu$K). Here we quote the \emph{rms} spread among the traps with the estimated contribution from statistical uncertainties per trap in parentheses.
 
\subsection{Magnetic field variations}
The precise values of the magnetic field strengths at the trap minima are also determined using rf spectroscopy performed after the evaporation sweep. We find site-to-site variations of $\Delta B=46(5)$~mG \emph{rms}, which appear randomly distributed over the lattice, but with some long-range correlations observed on the $100~\mu$m length scale. Compared with the total field produced by the film the figure of merit for trap variations is $\Delta B/B_y=2\times10^{-3}$ for a trap height of 10~$\mu$m. This value compares favourably with previous studies of permanent magnet atom chips for trap heights between 50~$\mu$m and 100~$\mu$m~\cite{SinRetHin05,WhiHalSid07}, but is a factor of $\gtrsim10$ larger than experiments using thin Au wires at comparable surface distances~\cite{TreGarBou07}, and approximately 100 times larger than for broad wires~\cite{KruAndSch07}. The precise origin of the variations is unknown, but could be associated with small imperfections in the patterning process or slight demagnetisation of the magnetic film during vacuum bakeout. We note that the requirements on the homogeneity of the applied bias fields are not very stringent ($\sim10^{-3}$ over 100~$\mu$m) and are easily satisfied by the external Helmholtz configuration coils. While inhomogeneity typically causes fragmentation of atom clouds in elongated trapping potentials, in our two-dimensional array each trapped cloud is much smaller than the characteristic length scale of the variations and experiences a single harmonic potential well. Furthermore, the small site-to-site variations affect our prospective qubit states equally (minimising dephasing~\cite{TreHomRei04}), and any residual effects could be precisely calibrated using site-resolved rf spectroscopy. 

\subsection{Density dependent decay}
During radio frequency evaporation, we find high cooling rates and high evaporation efficiencies ($d[\ln T]/d[\ln N]$ between 1.2 and 1.7), attributed to the extremely high elastic collision rates present in our stiff traps (between $2500-5000$~s$^{-1}$), despite the relatively small number of atoms per array site. At low temperatures ($<5~\mu$K), we find rf-spectroscopy becomes unreliable due to unavoidable spectral broadening caused by a combination of power-broadening, rapid decay from the untrapped state and depletion. Additionally, the influence of the rf-probe on the atomic energy distribution cannot be neglected at these temperatures, where collision rates are high, resulting in the redistribution of energy during the spectroscopy pulse. Instead, towards the end of the evaporation sweep we observe high decay rates ($\gamma>20$~s$^{-1}$) due to three-body recombination loss~\cite{BurGhrWie97,SodGueDal99}, which are accurately measured to characterise the array (figure~\ref{fig:evaporation}b). 

A collection of decay curves is obtained by holding the atoms for a variable duration (one value per experimental cycle) before taking an absorption image and integrating the number of atoms in each site. During the hold time we apply a radio-frequency ``shield'' tuned to the final evaporation frequency $\nu_f$ to limit the trap depth and prevent heating. Three-body decay curves are then modelled with the following differential equation:
\begin{eqnarray}
\dot{N}=-L\bigl{[}&{\textstyle \int} n_0^3d^3r+9{\textstyle \int} n_0^2n_{th}d^3r
&+18{\textstyle \int} n_{th}^2n_0d^3r+6{\textstyle \int} n_{th}^3d^3r\bigr{]}-N/\tau
\label{eq:loss}\end{eqnarray}%
where we have taken into account wavefunction symmetrisation (Ref.~\cite{KagSviShl85}), $L=1.8\times10^{-29}$~cm$^{6}$s$^{-1}$ is the three-body decay coefficient taken from Ref.~\cite{SodGueDal99} and $\tau=2~s$ is the $e^{-1}$ decay time for one-body (background) loss calibrated against dilute clouds trapped within the array. The atomic density distributions for a thermal fraction, $n_{\rm th}(r)$ and a Bose-condensed fraction, $n_0(r)$ for a given $N$ and $T$ are obtained using a self-consistent mean-field model. The model accounts for quantum statistics and the effect of interactions for both the thermal and condensed components but neglects the kinetic energy of the condensed fraction via the Thomas-Fermi approximation. We approximate the trap potential as isotropic $V(r)=m\bar{\omega}^2 r^2/2$ with mean trap frequency $\bar{\omega}=(\omega_\parallel\omega_\perp^2)^{1/3}$, and iteratively solve the following set of equations~\cite{GerThyAsp04}:
\begin{eqnarray}
  n_0(r)  =  \max\left[ \frac{1}{g}\left(\mu-V(r)-2g\; n_{\rm th}(r)\right),0 \right] \\
  n_{\rm th}(r)  =  \frac{1}{\Lambda^3}\;g_{3/2}\left[ e^{\left[\mu-V_{\rm eff}(r)\right] /k_B T}\right] \\
  V_{\rm eff}(r)  =  V(r)+2g\; (n_{\rm th}(r)+ n_{0}(r))
\end{eqnarray}
with $N=\int (n_0+n_{th}) d^3r$ and using the standard definitions: $g$ is the coupling strength, $\Lambda$ is the thermal de Broglie wavelength, $g_{n}[z]$ is the polylogarithm function and $\mu$ is the chemical potential. The numerical solutions are then used to predict the spectral energy distribution and three-body decay rate for fitting data as a function of $N$ and $T$ over the full range probed in our experiments. The solutions are also used to numerically determine the line $n(0)\Lambda^3=2.612$ in figure~\ref{fig:evaporation}c. For a more complete description of these models see ref.~\cite{GerThyAsp04}. 

We assume that thermal equilibrium is maintained due to the high elastic collision rate and we treat the cloud temperature as a constant parameter during decay. Fitting the numerical solution of equation~\eref{eq:loss} to each decay curve provides the initial $N$ and $T$. For all sets of experimental parameters the three-body decay curves are measured between 6 and 15 times and fit independently to reduce noise and estimate statistical uncertainties. For the intermediate temperatures ($\nu_f$=8300~kHz and $\nu_f=8500$~kHz datasets) we compare the results of both three-body decay and rf-spectroscopy measurements to calibrate systematic uncertainties such as the effective absorption cross-section used to determine $N$ from absorption images.

 The measurements taken after the evaporation sweep indicate peak atomic densities $>\!1\times10^{15}~$cm$^{-3}$ and temperatures around $\langle T\rangle=\nobreak2.0\pm0.3~\mu$K (blue-circles in figure~\ref{fig:evaporation}c) corresponding to 2.6 and 7.1 mean vibrational quanta in the transverse and axial modes respectively. From this we infer that each cloud is smaller than the optical wavelength (Lamb-Dicke regime), with \emph{rms} position spreads of $0.17\pm0.01~\mu$m and $0.4\pm0.03~\mu$m. Our data is consistent with $\sim$70 of the 156 analysed clouds having a phase-space-density ($\rho\!>\!\rho_0\!\simeq\!2.612$) corresponding to the onset of Bose-Einstein condensation. However, of these clouds, we infer an average condensate fraction of less than 0.1. Around this point the cooling efficiency sharply decreases due to increased three-body loss for partially condensed clouds, where preferential loss of atoms from the dense, Bose-condensed components during evaporation limits the final temperatures to around $T_c$ (solid line in figure~\ref{fig:evaporation}c). The observations are supported by our model which predicts three-body decay times, $\gamma^{-1}$, of around 20~ms (7~ms), for $N_{BEC}=50$ (200) respectively. This effect can be circumvented by using the weaker-confined $|F\!\!=\!\!1,m_F\!\!=\!\!-1\rangle$ state which also has a lower three-body-decay coefficient~\cite{BurGhrWie97}. Alternatively, three-body decay could be used to our advantage to produce a well defined number of atoms in each microtrap and to approach the few-atom regime for quantum information processing.

\begin{figure}[t] \centering\includegraphics[width=0.5\columnwidth]{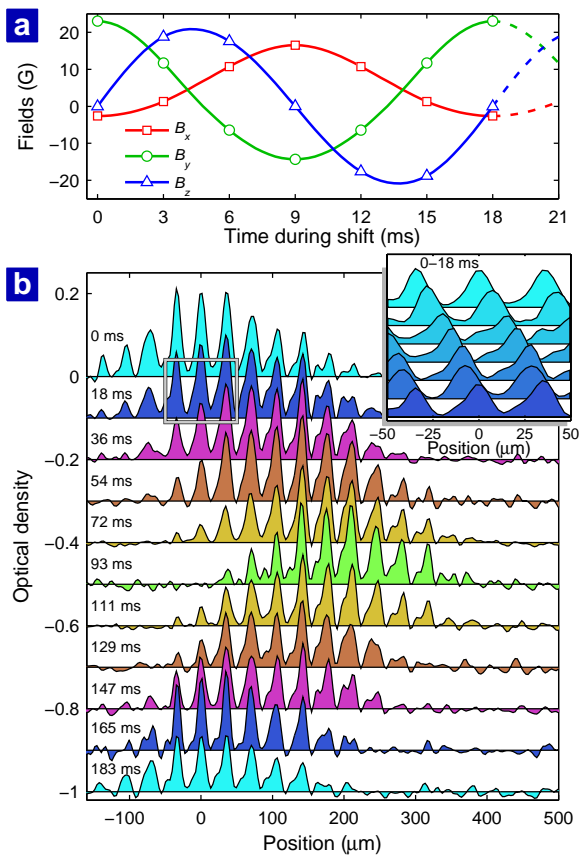}
\caption{{\bf Transport of atoms over the chip surface: atomic shift register.} An external homogeneous magnetic field with rotating orientation smoothly shuttles the array of atom clouds in one direction and back. {\bf (a)} Calculated optimal time-sequence for the external magnetic fields $B_x$ (red), $B_y$ (green) and $B_z$ (blue) applied to carry-out a single cycle shift operation. The field values shown as open symbols are applied in the experiment and interpolated by linear field ramps. {\bf (b)} Line-scans of optical density images along the shifting direction (arrow indicated in figure~\ref{fig:array}). Data for an integer number of shift cycles are shown, each with a vertical offset of -0.1 for clarity. A total of five shifts, each covering one lattice period, and five in the reverse direction to the original position are shown. The full sequence takes 183~ms and the atoms are transported over a distance of $360~\mu$m. The inset shows intermediate steps within the first shift cycle for each 3~ms interval.}
\label{fig:shiftregister}
\end{figure}

\section{Atomic shift register}
In addition to trapping and cooling the atom clouds we have transported the atoms using the atom chip as an atomic shift register. This parallelised version of atomic~\cite{HanHomRei01,KruLuoSch03} and molecular~\cite{MeeBetMei08} conveyor belts is the cold-atom analog of an electronic CCD (charge-coupled device). The atoms are trapped in local minima of the total magnetic field strength, which is the superposition of fields from the chip and the external field. Thus the position of field minima depends also on the external field orientation. By applying time-varying, external uniform fields in a cyclic fashion we can move the minima across the chip surface. Rotating the external magnetic field continuously displaces the traps, shifting the array along one lattice direction as shown in figure~\ref{fig:shiftregister}. We have computed the optimised field components (figure~\ref{fig:shiftregister}a) required to keep the traps at a fixed height of $10~\mu$m and with the field minima at a constant $3.23~$G. The shifting direction is reversed by simply reversing the order of the applied fields. Details of this calculation, as well as restrictions to where one can or cannot move the atoms can be found in Ref.~\cite{GerSpr06}. 

In the experiment we first cool the atom clouds to a temperature of $\sim\!10~\mu$K and then linearly ramp between six settings of the external field per period, in $6\times3$~ms intervals. We investigate the operation of our atomic shift register using thermal clouds which provide a good signal to noise in a regime where interactions can be neglected and three-body loss is reduced. By repeating the shifting sequence we have transported the atoms over a round-trip distance of $360~\mu$m, with five shifts in the $(2,-3)$ direction (figure~\ref{fig:array}) followed by five reverse shifts (figure~\ref{fig:shiftregister}b). During the experiment we observe about 40\% atom loss. This is consistent with expectations for three-body loss over the 183~ms experiment duration. We have also measured the effect of the shift register on the temperature of the clouds using rf-spectroscopy. In the experiment we keep the total trapping time after the evaporation sweep constant, but in the first measurement we transport the atoms over 5 lattice periods, while in the second measurement they remain trapped in their original positions. We measure mean temperatures of 10.9$\pm 0.7~\mu$K and 11.5$\pm 0.6~\mu$K with and without transport respectively, consistent with no observed additional heating. The demonstrated ability to smoothly shuttle atoms across the chip is an important ingredient for a scalable architecture for quantum computing and is an inherent feature of our magnetic-film atom chip.

\section{Single site addressing}
\begin{figure}[t] \centering \includegraphics[width=0.7\columnwidth]{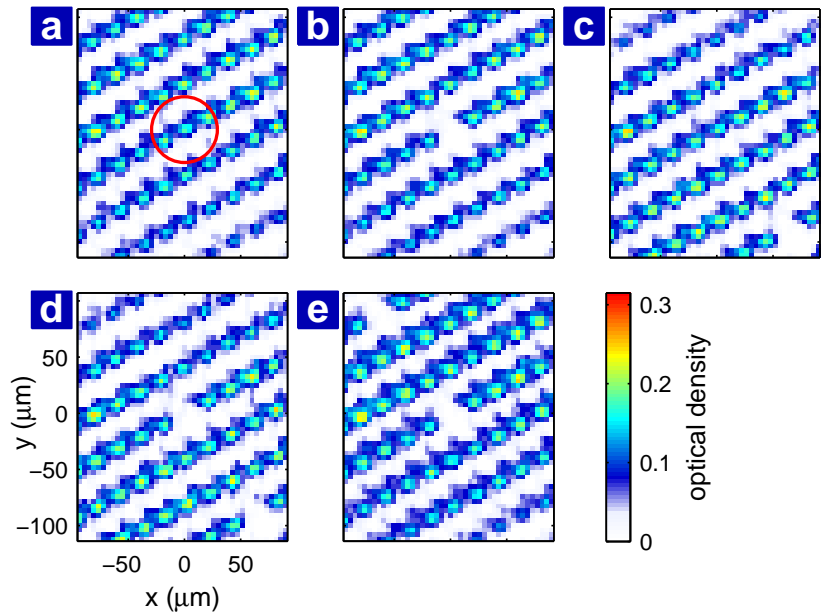}
\caption{{\bf Single site addressability and transport of an empty site.} Sequence of images demonstrating single site addressability using a focused laser pulse to empty a selected trap and transport using the atomic shift register. {\bf (a)} Atoms are loaded to the lattice at a temperature of $\sim10~\mu$K. The center of the red circle indicates the position of the focused addressing laser spot. {\bf (b)} Optical pumping light focused on one lattice site is pulsed on for 1~ms to drive transitions to untrapped Zeeman states, completely emptying the microtrap. The effect of the addressing pulse on neighbouring traps is negligible. {\bf (c)} All the traps along with the emptied site are shifted by three lattice periods to a new position (bottom right). {\bf (d)} The addressing laser is pulsed on again to empty a second trap, and finally {\bf (e)} the array is shifted back to the original position.}
\label{fig:addressing}
\end{figure}

Finally we show the addressable nature of our microtrap array by selectively emptying a single site near the center of the array using a focused addressing laser pulse. The laser is frequency stabilized to the $F=2\rightarrow F'=2$ optical pumping transition of the D2 line which has a high probability of changing the spin of the atoms to an untrapped Zeeman state (0.5 per scattered photon). The addressing laser is applied by imaging of the output facet of a single mode optical fiber to the atom chip. We make double use of the optical imaging system to focus the light: positioning the fiber in a second image plane (at the same distance as the CCD camera from the chip) created with a beamsplitter. We estimate a focused spot corresponding to a $\sim7~\mu$m Airy disc radius at the atom chip, which is highly homogeneous over the extent of a single cloud but is small compared to the distance between neighbouring traps. The spot is then easily aligned on any single array site by controlling the fiber position with an x-y translation stage and imaging the back-reflected light on the CCD camera for position reference.

Figure~\ref{fig:addressing}a shows part of an absorption image taken of the array after an evaporative cooling sweep to $T\approx10~\mu$K. The focused addressing laser is aligned on the site at the center of the image as indicated by the red circle. The addressing laser is then pulsed on for 1~ms to completely empty this trap of atoms (figure~\ref{fig:addressing}b). We do not measure any atom loss or heating caused by the addressing pulse on neighboring atom clouds, even for pulse durations as long as 10~ms, indicating excess light incident on neighbouring traps is negligible. We then demonstrate how optical addressing can be combined with the atomic shift register in order to write patterns to the array. Figure~\ref{fig:addressing}c shows the lattice after emptying the central microtrap and shifting the lattice by three periods (lower-right). Subsequently we can pulse on the addressing laser to empty another trap (figure~\ref{fig:addressing}d). Finally the lattice is shifted again three periods in the reverse direction back to the original position. The addressed traps remain empty for the full duration of the experiment of 126~ms. The same techniques could be used to create and readout spin coherence between quantum states of atoms in a single trap. These experiments clearly demonstrate local manipulation of individual atom clouds in our array and point to the possibility of encoding and shifting quantum information. It is also possible to use focused resonant imaging light for the addressing laser, to detect atoms in one trap without affecting neighbouring traps of the array. Focused lasers are particularly powerful for addressing two-dimensional arrays, especially for our system where atoms are confined to regions much smaller than the distance between traps. Straightforward modifications to our setup using acousto-optical or spatial light modulators to control the position of the addressing laser will allow for sequential or simultaneous writing to the quantum state of atoms trapped in each microtrap. 

We expect this system to be the ideal platform for studying controlled many particle entanglement and quantum information processing with neutral atoms on an atom chip. Specifically, qubits could be encoded in two magnetically trappable states $|F=1,m_F=-1\rangle$ and $|F=2,m_F=+1\rangle$ of the $^{87}$Rb electronic ground state. With focused Raman lasers it is possible to coherently drive transitions between logical states in targeted traps. Focused lasers could also be used to state-selectively drive atoms to highly excited Rydberg states. In these small traps, the dipole blockade mechanism~\cite{JakCirLuk00,LukFleZol01} would limit the process to a single Rydberg excitation per trap and could mediate interactions on demand over distances of tens of micrometers, between neighbouring traps~\cite{DitKoeHeu08,UrbJohSaf08}. If required, single atom detection sensitivity could be achieved through subsequent field ionisation of the excited atoms. Alternatively, entangled mesoscopic ensembles of atoms could be produced and used for quantum information processing  using Rydberg excitations~\cite{LukFleZol01,MulLesZol08}. Future studies will be aimed at investigating the effect of magnetic and electric field noise near the chip surface on the lifetime and phase coherence of ground state and excited atoms. 

\section{Conclusion}
To conclude, we have reported experiments on a two-dimensional array of ultracold atom clouds prepared near the surface of a magnetic film atom chip. This promising new architecture for scalable quantum information processing combines optically-resolved trap separations with tight magnetic confinement. We have loaded atoms into more than 500 array sites and performed rf evaporative cooling of these atom clouds to the quantum degeneracy temperature. In the process we have demonstrated preparation, cooling, detection and analysis of hundreds of atom clouds in parallel; highlighting magnetic microtrap arrays as a unique experimental test bed for studying ultracold atomic systems and quantum matter. Using existing microfabrication technology we anticipate further scaling, to lattice dimensions of a few micrometers, in order to accommodate thousands of individual atomic qubits. An atomic shift register is used for parallel transport of atoms across the chip surface, allowing the separation of storage, interaction and readout areas of a future, fully integrated device. Finally, we demonstrate local manipulation of trapped atoms by selectively emptying individual traps using a focused optical pumping laser.

\ack{ We would like to thank M. S. Golden, H. B. van Linden van den Heuvell and N. J. van Druten for helpful
discussions and careful reading of the manuscript. We thank J.-U. Thiele (Hitachi San Jose Research Center) and the group of J.~B. Goedkoop (U. of Amsterdam) for developing the FePt films, and C. Ockeloen and B. Rem for work on the optical imaging setup. The chips were patterned using the  facilities of the Amsterdam {\it nano}Center. This work is part of the research programme of the `Stichting voor Fundamenteel Onderzoek der Materie (FOM)', which is financially supported by the `Nederlandse Organisatie voor Wetenschappelijk Onderzoek (NWO)'. It was also supported by the EU under contract MRTN-CT-2003-505032. SW acknowledges support from a Marie-Curie individual fellowship (grant number PIIF-GA-2008-220794).}

\section*{References}
\bibliography{swhitlock08NJP}

\end{document}